\documentclass[a4paper]{article}
\usepackage{ISCSLP2026}
\usepackage{ifthen}
\usepackage{amsmath,graphicx,hyperref}

\usepackage{cite}
\usepackage{newtxmath}
\usepackage{booktabs,multirow}
\newboolean{blind}
\setboolean{blind}{false}

\newcommand{\best}[1]{\textbf{#1}}
\usepackage{comment}

\title{A Large-Scale Probing Analysis of Speaker-Specific Attributes in Self-Supervised Speech Representations}
\name{
	\ifthenelse{\boolean{blind}}{Anonymous to ISCSLP}
	{{Aemon Yat Fei} {Chiu}$^{\ddagger}$, {Kei Ching} {Fung}$^{\ddagger}$, {Roger Tsz Yeung} {Li}$^{\ddagger}$, {Jingyu} {Li}$^{\dagger}$, {Tan} {Lee}$^{\diamond,\ddagger}$}
}
\address{
  \ifthenelse{\boolean{blind}}{Anonymous to ISCSLP}
  {
  	$^{\ddagger}$Department of Electronic Engineering, The Chinese University of Hong Kong, Hong Kong \\
    $^{\dagger}$Independent Researcher \\
    $^{\diamond}$School of Data Science, The Chinese University of Hong Kong, Shenzhen, China
  }
}

\email{
	\ifthenelse{\boolean{blind}}{Anonymous to ISCSLP}
	{\{aemon.yf.chiu,pacokcfung,rogerli,lijingyu0125\}@link.cuhk.edu.hk, tanlee@ee.cuhk.edu.hk}
}

\begin{document}

\maketitle
\begin{abstract}
  Enhancing explainability in speech self-supervised learning (SSL) is important for understanding and effectively utilising speech SSL representations. This study conducts a large-scale layer-wise probing analysis of 11 speech SSL models, examining speaker identity together with acoustic, prosodic, and paralinguistic attributes. The results confirm a general hierarchy wherein initial layers encode fundamental acoustics and middle layers synthesise abstract traits. The consensus that final layers purely abstract linguistic content is challenged, as larger models unexpectedly recover speaker-discriminative information in their deep layers. Furthermore, the intermediate representations of speech SSL models are found to provide stronger probing performance for several attributes than specialised speaker embeddings. These insights give a systematic empirical characterisation of speaker-related information in SSL models, providing guidelines for selecting representations for downstream tasks.
\end{abstract}
\noindent\textbf{Index Terms}: speech representation learning, speaker-specific attribute, probing, explainable AI.

\section{Introduction}
\label{sec:intro}

\begin{figure}[t]
  \centering
  \includegraphics[width=0.95\linewidth]{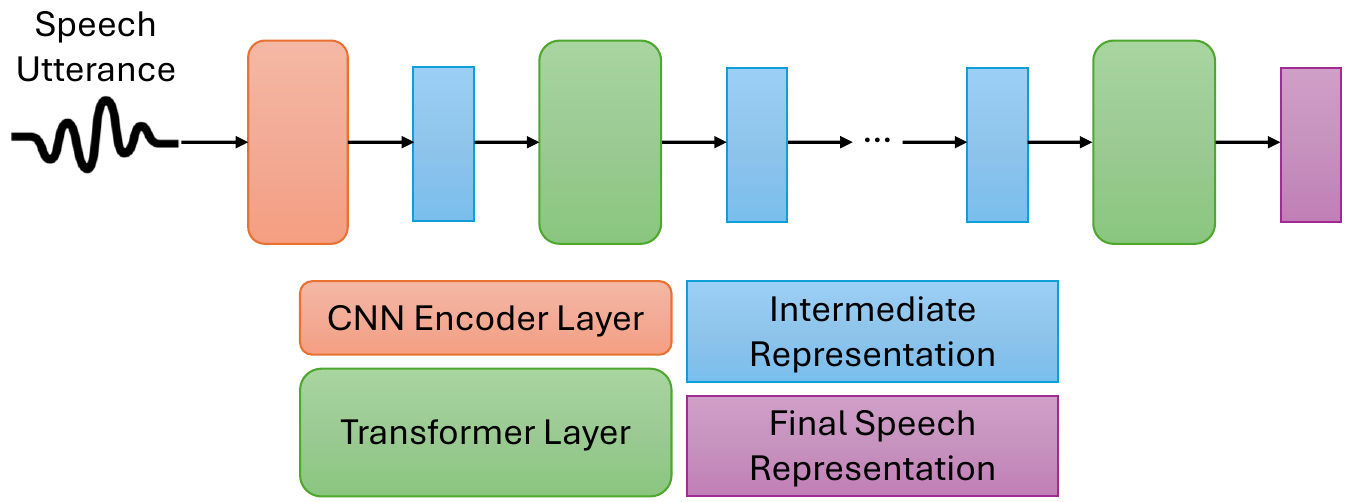}
  \vspace{-0.75em}
  \caption{A schematic diagram of typical speech SSL models~\cite{wav2vec2, hubert, unispeech-sat, wavlm}.}
\label{fig:ssl}
\vspace{-1.85em}
\end{figure}

Speech representation learning has rapidly advanced in recent years~\cite{ssl, yang2024taslp, wang2024taslp}, driven by the success of large-scale Transformer-based \cite{attention} speech self-supervised learning (SSL) models such as Wav2vec 2.0~\cite{wav2vec2}, HuBERT~\cite{hubert}, UniSpeech-SAT~\cite{unispeech-sat}, and WavLM~\cite{wavlm}. These models have transformed speech representation learning by extracting rich and hierarchical representations from massive amounts of unlabelled audio, achieving state-of-the-art performance on a wide range of downstream speech processing tasks. Typical speech SSL models generally comprise a convolutional neural network (CNN) encoder and a series of Transformer layers, as illustrated in Figure~\ref{fig:ssl}~\cite{wav2vec2, hubert, unispeech-sat, wavlm}. However, like other models based on deep neural networks (DNN), these speech SSL models essentially operate as complex black boxes. Understanding their internal mechanics is significant for their effective utilisation. It is accepted that the lower layers of these models capture acoustic details, while the upper layers abstract linguistic content~\cite{ssl, yang2024taslp}. However, this high-level view does not fully capture the complexity of the information encoded, particularly regarding the varied nature of a speaker's voice.

Human communication is not a direct mapping from sound to words. Rather, it is a complex informational signal that can be decomposed into distinct streams. Prior work in phonetics~\cite{semiotic} and modern text-to-speech synthesis (TTS)~\cite{SpeechTripleNet,gengembre24_interspeech} distinguishes between linguistic content, timbre, and prosody. Understanding how speech SSL models represent these distinct attributes is crucial for improving interpretability and developing more reliable downstream applications.

This study serves as a comprehensive explainability analysis and is motivated by the following research questions: \textbf{how do speech SSL models represent different categories of speaker-specific attributes across their layers, and how do various model families and scales differ in encoding these attributes?} While previous studies have investigated various acoustic and linguistic properties encoded in deep speaker embeddings \cite{wang17g_Interspeech, raj19asru} and speech SSL representations \cite{pasad2021asru, pasad2023icassp, choi24b_interspeech, singla2022icdmw, Ashihara}, the present study provides a unified empirical characterisation of a curated set of speaker-specific attributes across 11 speech SSL models spanning four model families and multiple scales. By framing the analysis of continuous attributes as a classification task with discretised labels, a direct comparison across models and layers is introduced. This approach facilitates an understanding of how these large models represent different acoustic and stylistic elements of speech, giving practical implications for model selection in downstream tasks.

\section{Related Work}
\label{sec:review}

\subsection{Probing single-vector deep speaker embeddings}
\label{sec:review_dse}

The use of simple classifiers to ``probe'' the information contained in learnt representations was established in foundational studies of deep speaker embeddings (e.g., $i$-vectors, $d$-vectors, and $s$-vectors), examining whether they capture fine-grained speaker attributes like gender and speaking rate~\cite{wang17g_Interspeech}. Raj et al. \cite{raj19asru} extended this analysis to $x$-vectors, where they also analysed the model's robustness to data augmentation.

While prior work demonstrated that speaker embeddings trained for speaker verification (SV) also captured many other attributes, including speaking rate, gender, and linguistic content, the present study extends the probing paradigm to Transformer-based speech SSL models, focusing on the evolution of speaker-related information across layers.

\begin{figure}[t]
  \centering
  \includegraphics[width=\linewidth]{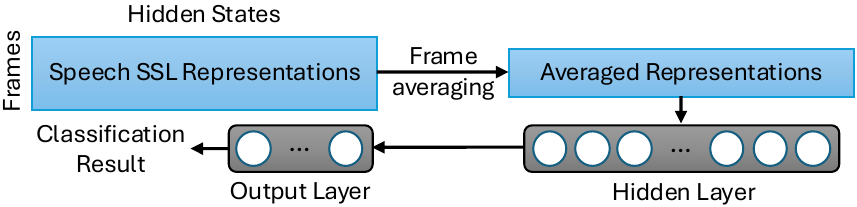}
  \vspace{-2em}
  \caption{The overall design of the probing task for speech SSL representations.}
\label{fig:probing}
\vspace{-1.85em}
\end{figure}

\subsection{Layer-wise analysis of speech SSL models}
\label{sec:review_ssl}

Recent research has provided extensive analyses of speech SSL models, employing a range of methodologies and focusing on different aspects of speech. Canonical correlation analysis was used in \cite{pasad2021asru, pasad2023icassp} to map the layer-wise distribution of acoustic, phonetic, and word-level information, revealing how it is shaped by a model's pre-training objective. A subsequent study \cite{choi24b_interspeech} used direct similarity metrics to show that speech SSL representations are more sensitive to phonetic similarity than semantic similarity.

Singla et al. \cite{singla2022icdmw} comprehensively probed an extensive set of 46 features related to language delivery and structure, creating a detailed inventory of what is learnt in each layer. Ashihara et al. \cite{Ashihara} used probing tasks to compare the capabilities of general-purpose speech SSL models against specialised speaker SSL models, finding that the former retain more linguistic information while the latter are more capable of capturing speaker identity. The present study is distinct from the literature in its specific analytical focus by employing a concisely grouped set of speaker-specific acoustic and prosodic attributes derived from phonetic literature and systematically comparing 11 speech SSL models across multiple families and scales.

\section{Experimental Settings}

The probing method is based on the principle that information readily accessible in a representation should enable a relatively simple classifier to predict the corresponding attribute \cite{wang17g_Interspeech}. The probing pipeline is constructed as illustrated in Figure \ref{fig:probing}.

\subsection{Probing network}

Following the classifier designed in~ \cite{wang17g_Interspeech,raj19asru}, a simple classification network, i.e., a multilayer perceptron (MLP), is utilised as the probing model. The MLP has a single hidden layer with $500$ nodes followed by a ReLU activation. The input size of the MLP corresponds to the dimensionality of the hidden representation in the speech SSL model, as these representations are frame-averaged before being fed into the MLP.

\subsection{Dataset}
\label{sec:dataset}

The TextrolSpeech dataset \cite{ji2024icassp} is used for the training and test datasets in the probing tasks. TextrolSpeech is a large-scale dataset comprising over 330 hours of English speech data with rich labels for controllable TTS. It integrates LibriTTS \cite{zen19_interspeech}, VCTK \cite{vctk}, and emotional TTS datasets, including ESD \cite{ZHOU20221ESD}, TESS \cite{tess}, MEAD \cite{mead}, SAVEE \cite{savee}, and MESS \cite{mess}. These last five emotional datasets make up 16\% of the corpus and contain labels for gender, pitch, tempo, energy, and emotion. LibriTTS and VCTK are also enriched with these labels, although their emotion labels are assigned as ``Neutral''. A total of 1,327 speakers are included in the dataset, which is partitioned into training and test sets. In this study, 90\% of the original TextrolSpeech training set is randomly sampled at the utterance level to form the training set, totalling 212,400 utterances, while the remaining 10\% (23,603 samples) forms the validation set. Following training and validation, the MLP is evaluated on a test set containing 200 utterances, which is identical to the test set in the original TextrolSpeech dataset.

\begin{figure}[t]
  \centering
  \includegraphics[width=0.95\linewidth]{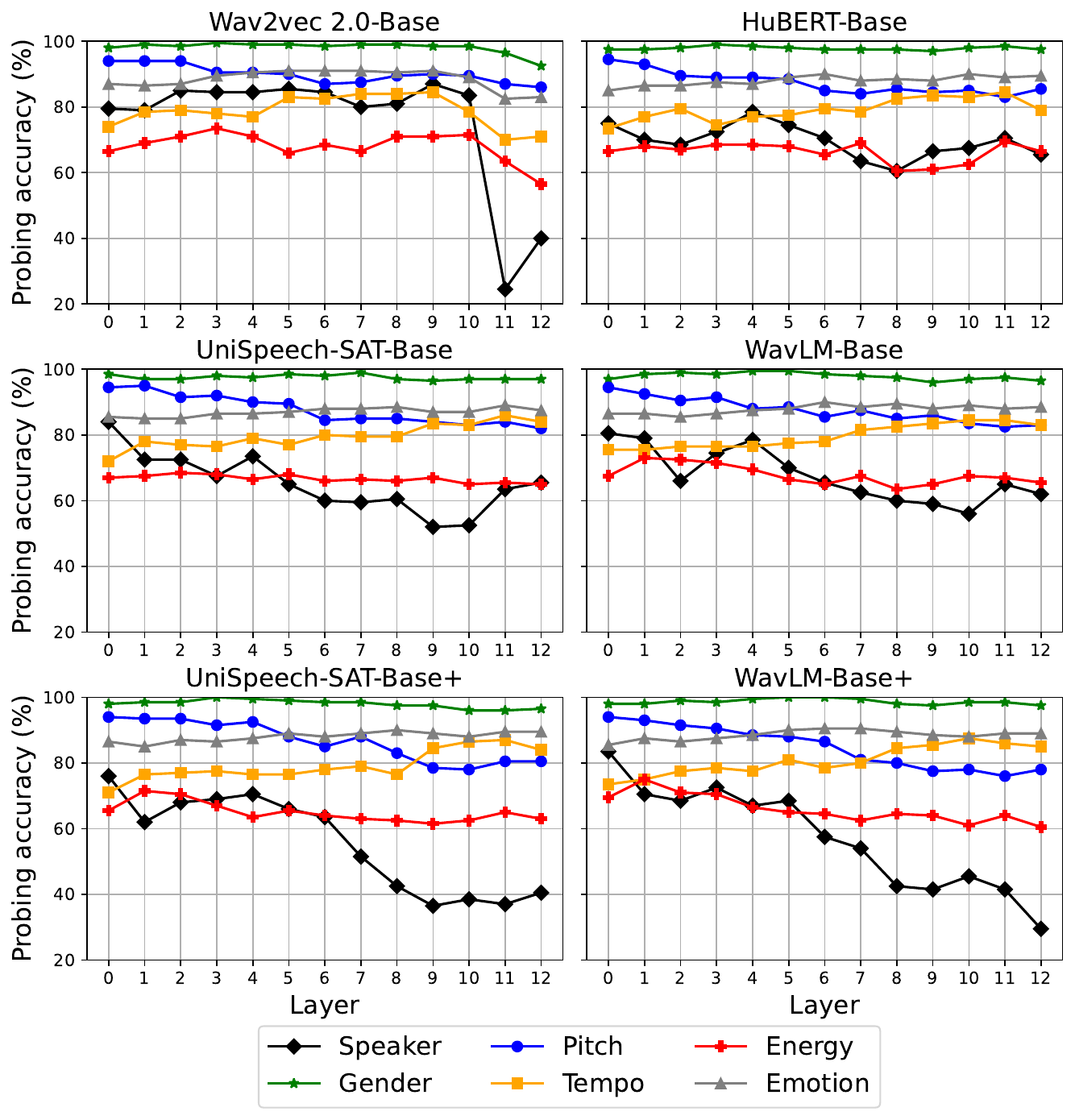}
  \vspace{-1.125em}
  \caption{Probing accuracies for the \textit{Base} and \textit{Base-Plus} variants of speech SSL models on the test set.}
\label{fig:base}
\vspace{-1.85em}
\end{figure}

\begin{figure*}[t!]
  \centering
  \includegraphics[width=0.95\linewidth]{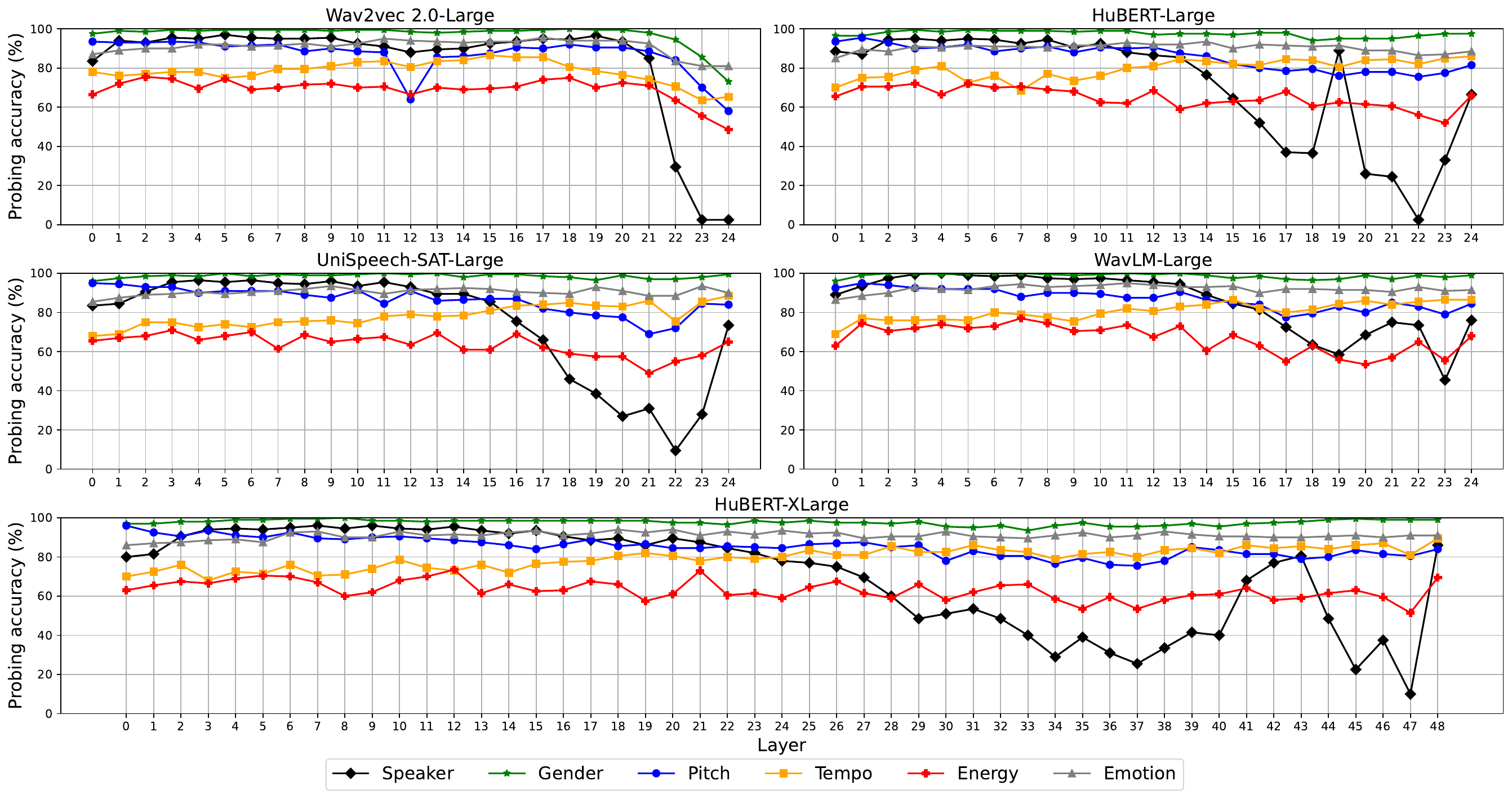}
  \vspace{-1.25em}
  \caption{Probing accuracies for the \textit{Large} and \textit{XLarge} variants of speech SSL models on the test set.}
\label{fig:large}
\vspace{-2em}
\end{figure*}

\subsection{Probed speaker-specific attributes}

\textbf{Speaker} identity and five speaker-specific features, categorised into three categories, are probed in this study: \textbf{Gender}, acting as a proxy for stable and long-term timbre quality features closely related to \textit{acoustic} characteristics; \textbf{Pitch}, \textbf{Tempo}, and \textbf{Energy}, representing three core utterance-level attributes of \textit{prosody}; and \textbf{Emotion}, representing an affective state and \textit{paralinguistic} trait. In this study, \textbf{Gender} and \textbf{Emotion} are manually labelled. \textbf{Pitch}, \textbf{Tempo}, and \textbf{Energy} levels are derived from the dataset using acoustic processing tools, and each audio sample is labelled as ``High'', ``Normal'', or ``Low'' based on the overall distribution within the TextrolSpeech dataset~\cite{ji2024icassp}. \textbf{Emotion} labels include ``Happy'', ``Contempt'', ``Disgusted'', ``Angry'', ``Fear'', ``Surprised'', ``Sad'', and ``Neutral''.

\subsection{Models under investigation}

The intermediate representations from 11 pre-trained speech SSL models, sourced from \textit{Hugging Face}, are analysed. The selection covers four major model families at multiple scales to ensure broad coverage across architectures and scales:

\begin{itemize}
    \item \textbf{Wav2vec 2.0}: Wav2vec 2.0-\textit{Base}, \textit{Large}.
    \item \textbf{HuBERT}: HuBERT-\textit{Base}, \textit{Large}, \textit{XLarge}.
    \item \textbf{UniSpeech-SAT}: UniSpeech-SAT-\textit{Base}, \textit{Base-Plus}, \textit{Large}.
    \item \textbf{WavLM}: WavLM-\textit{Base}, \textit{Base-Plus}, \textit{Large}.
\end{itemize}

The \textit{Base} and \textit{Base-Plus} models have 12 Transformer layers, outputting 768-dimensional (\textit{d}) intermediate representations. \textit{Large} variants, comprising 24 Transformer layers, produce 1,024-\textit{d} representations, while the \textit{XLarge} model has 48 Transformer layers and produces 1,280-\textit{d} representations.

\subsection{Training settings}

Following \cite{raj19asru}, cross-entropy loss is employed as the loss function for the classification tasks. The Adam optimiser is utilised with a learning rate of $0.001$. The batch size is set at $1,024$ and the model is trained for one epoch on the training set during each round. Mixed precision training is employed to accelerate the process. For each model and attribute, the layer with the highest validation accuracy is selected, and its corresponding test accuracy is reported.

\begin{table*}[t]
\centering
\caption{Validation accuracies (\%, denoted as \textbf{Val}), corresponding test accuracies (\%, denoted as \textbf{Test}), and the selected layer (denoted as \textbf{L}) for all configurations. For each setting, \textbf{L} is selected according to the best \textbf{Val}. Layer 0 refers to the CNN encoder layer.}
\vspace{-0.7em}
\resizebox{0.9875\linewidth}{!}{
\setlength{\tabcolsep}{3pt}

\begin{tabular}{@{}l *{6}{c c c}@{}}
\toprule
\multicolumn{1}{c}{\multirow{2}{*}{\textbf{Model}}} &
\multicolumn{3}{c}{\textbf{Speaker}} & \multicolumn{3}{c}{\textbf{Gender}} &
\multicolumn{3}{c}{\textbf{Pitch}}   & \multicolumn{3}{c}{\textbf{Tempo}}  &
\multicolumn{3}{c}{\textbf{Energy}}  & \multicolumn{3}{c}{\textbf{Emotion}} \\
\cmidrule(lr){2-4}
\cmidrule(lr){5-7}
\cmidrule(lr){8-10}
\cmidrule(lr){11-13}
\cmidrule(lr){14-16}
\cmidrule(lr){17-19}
& \textbf{L} & \textbf{Val} & \textbf{Test}
& \textbf{L} & \textbf{Val} & \textbf{Test}
& \textbf{L} & \textbf{Val} & \textbf{Test}
& \textbf{L} & \textbf{Val} & \textbf{Test}
& \textbf{L} & \textbf{Val} & \textbf{Test}
& \textbf{L} & \textbf{Val} & \textbf{Test} \\
\midrule

Wav2vec 2.0-\textit{Base}
& 3  & 84.3 & 84.5
& 3  & 99.1 & 99.5
& 1  & 93.2 & 94.0
& 8  & 85.3 & 84.0
& 3  & 69.5 & 73.5
& 7  & 93.2 & 91.0 \\

HuBERT-\textit{Base}
& 0  & 76.8 & 75.0
& 4  & 98.8 & 98.5
& 1  & 92.6 & 93.0
& 9  & 85.0 & 83.5
& 4  & 67.6 & 68.5
& 11 & 92.3 & 89.0 \\

UniSpeech-SAT-\textit{Base}
& 0  & 81.3 & 84.0
& 4  & 98.9 & 97.5
& 0  & 92.5 & \best{94.5}
& 10 & 84.8 & 83.0
& 1  & 67.4 & 67.5
& 8  & 92.0 & 88.5 \\

UniSpeech-SAT-\textit{Base-Plus}
& 0  & 77.9 & 76.0
& 3  & 98.9 & \best{100.0}
& 0  & 92.5 & 94.0
& 11 & 86.5 & 87.0
& 3  & 67.8 & 67.0
& 7  & 92.3 & 89.0 \\

WavLM-\textit{Base}
& 0  & 81.8 & 80.5
& 4  & 99.0 & 99.5
& 1  & 92.2 & 92.5
& 10 & 85.8 & 84.5
& 0  & 69.9 & 67.5
& 7  & 91.7 & 88.5 \\

WavLM-\textit{Base-Plus}
& 0  & 83.6 & 83.5
& 3  & 99.0 & 98.5
& 0  & 92.3 & 94.0
& 10 & 87.6 & 87.5
& 1  & 72.0 & \best{75.0}
& 7  & 92.1 & 90.5 \\

\midrule

Wav2vec 2.0-\textit{Large}
& 19 & 96.0 & 96.5
& 19 & \best{99.4} & 99.5
& 1  & \best{94.1} & 93.0
& 15 & 86.4 & 86.5
& 19 & \best{73.4} & 70.0
& 15 & 95.0 & 93.5 \\

HuBERT-\textit{Large}
& 3  & 95.2 & 95.0
& 6  & 99.1 & 99.0
& 3  & 92.7 & 90.0
& 24 & 85.7 & 86.0
& 5  & 69.3 & 72.0
& 13 & 94.7 & 92.0 \\

UniSpeech-SAT-\textit{Large}
& 4  & 96.5 & 96.5
& 4  & 99.2 & 98.5
& 2  & 92.9 & 93.0
& 24 & 86.4 & 88.5
& 3  & 68.4 & 71.0
& 9  & 94.5 & 93.5 \\

WavLM-\textit{Large}
& 3  & \best{98.7} & \best{99.5}
& 4  & \best{99.4} & 99.5
& 3  & 93.6 & 92.5
& 20 & 86.8 & 86.0
& 3  & 72.7 & 72.0
& 10 & \best{95.7} & \best{94.0} \\

\midrule

HuBERT-\textit{XLarge}
& 6  & 95.6 & 95.0
& 7  & 99.2 & 99.5
& 3  & 92.8 & 93.5
& 48 & \best{87.7} & \best{89.5}
& 12 & 67.1 & 73.5
& 15 & 95.0 & 93.5 \\

\bottomrule
\end{tabular}
}
\vspace{-1.7em}
\label{tab:peak_results}
\end{table*}

\begin{table}[b!]
\vspace{-1.4em}
    \centering
    \caption{Probing accuracies ($\%$) of deep speaker embedding models on the validation (\textbf{Val}) and test (\textbf{Test}) sets.}
    \vspace{-0.7em}
    \label{tab:embed}
    \resizebox{\columnwidth}{!}{
    \begin{tabular}{ccccccc}
\toprule
\multirow{2}{*}{\textbf{Attribute}} &
\multirow{2}{*}{\textbf{Data}} &
E-TDNN & E-TDNN & ResNet & ResNet &
\multirow{2}{*}{CAM++} \\
& & 512 & 1024 & 34 & 293 & \\
\midrule
\midrule
\multirow{2}{*}{\textbf{Speaker}} & \textbf{Val} & 98.08 & 98.02 & \textbf{98.39} & 98.14 & 98.03 \\
                                   & \textbf{Test} & 98.50 & 98.50 & 98.00 & 98.50 & \textbf{99.00} \\
\cmidrule(r){1-2}\cmidrule(r){3-4}\cmidrule(r){5-6}\cmidrule(r){7-7}
\multirow{2}{*}{\textbf{Gender}} & \textbf{Val} & 99.41 & 99.42 & 99.53 & 99.48 & \textbf{99.71} \\
                                 & \textbf{Test} & \textbf{100.00} & \textbf{100.00} & \textbf{100.00} & \textbf{100.00} & 99.50 \\
\cmidrule(r){1-2}\cmidrule(r){3-4}\cmidrule(r){5-6}\cmidrule(r){7-7}
\multirow{2}{*}{\textbf{Pitch}} & \textbf{Val} & 85.70 & 83.71 & \textbf{86.31} & 84.04 & 86.12 \\
                                & \textbf{Test} & \textbf{86.50} & 81.00 & \textbf{86.50} & 79.00 & 84.50 \\
\cmidrule(r){1-2}\cmidrule(r){3-4}\cmidrule(r){5-6}\cmidrule(r){7-7}
\multirow{2}{*}{\textbf{Tempo}} & \textbf{Val} & 68.20 & 67.78 & 68.39 & 67.49 & \textbf{68.66} \\
                                & \textbf{Test} & 67.50 & 70.00 & \textbf{71.50} & 70.50 & 71.00 \\
\cmidrule(r){1-2}\cmidrule(r){3-4}\cmidrule(r){5-6}\cmidrule(r){7-7}
\multirow{2}{*}{\textbf{Energy}} & \textbf{Val} & 65.47 & 64.63 & 66.82 & 65.28 & \textbf{67.42} \\
                                 & \textbf{Test} & 63.50 & 63.50 & 66.00 & 66.50 & \textbf{67.00} \\
\cmidrule(r){1-2}\cmidrule(r){3-4}\cmidrule(r){5-6}\cmidrule(r){7-7}
\multirow{2}{*}{\textbf{Emotion}} & \textbf{Val} & 91.66 & 90.36 & 91.09 & 89.74 & \textbf{94.41} \\
                                  & \textbf{Test} & 88.50 & 88.00 & 87.00 & 84.50 & \textbf{93.00} \\
\bottomrule
\end{tabular}
    }
\end{table}

\section{Results and Analyses}

Figure~\ref{fig:base} illustrates the probing accuracy on the TextrolSpeech test data for the \textit{Base} and \textit{Base-Plus} speech SSL models \cite{wav2vec2, hubert, unispeech-sat, wavlm} across different layers, while Figure~\ref{fig:large} demonstrates this for the \textit{Large} and \textit{XLarge} variants. A summary of validation-selected peak layer with corresponding accuracies for each attribute given by each model is illustrated in Table~\ref{tab:peak_results}.

\subsection{Hierarchy of attribute encoding}
\label{sec:hierarchical}

Despite differences in model architectures and pre-training data, all tested models exhibit several broadly consistent but attribute-dependent layer-wise patterns, consistent with the general hierarchical organisation reported in previous studies~\cite{ssl, yang2024taslp}.

The initial layers of those models function as powerful acoustic feature extractors. This is evidenced by \textbf{Pitch}, which consistently achieves its peak probing accuracy in the early layers. \textbf{Energy} is also often highly accessible in relatively shallow layers, although its peak layer varies substantially across models. \textbf{Gender}, acting as a stable acoustic attribute, maintains high probing accuracy across all layers. The probing accuracy curves for \textbf{Emotion} are also relatively stable from the initial to the final layers, indicating that emotion-related information remains generally accessible across substantial network depth. The probing accuracies for \textbf{Speaker} are often high in the initial layers, although the optimal layer varies considerably across models.

\textbf{Tempo} exhibits a different layer-wise pattern from \textbf{Pitch}. Its peak layers occur substantially deeper in the network. This suggests that information associated with utterance-level speaking rate remains highly accessible after more acoustic information has been transformed by deeper layers.

\textbf{Speaker} identity exhibits more model-dependent behaviour. Speaker-discriminative information is highly accessible in early-to-middle representations for many models, but the pattern is not universally monotonic. For example, Wav2vec 2.0-\textit{Large} reaches its highest validation \textbf{Speaker} accuracy at Layer 19. Several \textit{Large} and \textit{XLarge} models also show a decrease in \textbf{Speaker} probing accuracy at intermediate depths followed by an increase in later layers. These results indicate that speaker-discriminative information can remain or again become strongly accessible in deep representations.

\subsection{Relationship between speaker identity and probed attributes}
\label{sec:deconstruction}

The probing results demonstrate that speaker-discriminative information is often highly accessible in the early-to-mid layers, where \textbf{Gender}, \textbf{Pitch}, and frequently \textbf{Energy} are also strongly represented. This correspondence indicates that speaker-discriminative information and several speaker-associated acoustic or prosodic attributes coexist within related regions of the representation hierarchy.

The probing accuracies for \textbf{Energy} are consistently lower than those for features such as \textbf{Gender} and \textbf{Pitch}. Although its accuracy also tends to peak in the early layers, its lower overall performance suggests that Energy is less readily accessible under the present probing protocol.

\subsection{Model-specific variations and the impact of scale}

While the general information hierarchy is broadly similar across model families, the probing results reveal certain variations among different model families and scales.

For speaker-related tasks, models with pre-training objectives targeting noisy and multi-speaker environments, such as WavLM~\cite{wavlm}, provide particularly strong speaker-discriminative representations. WavLM-\textit{Large} achieves particularly strong probing performance for \textbf{Speaker}.

However, analysing the models across different scales reveals an important trade-off. While larger model variants exhibit clear performance gains on complex high-level traits like \textbf{Speaker} and \textbf{Emotion}, improvements for fundamental acoustic and prosodic features are often marginal. This suggests that for downstream tasks involving fine-grained analysis of speaker-specific features, smaller and more efficient \textit{Base} or \textit{Base-Plus} models may offer sufficient performance and an improved performance-to-cost ratio. Conversely, tasks requiring high-level speaker or emotion modelling tend to benefit from larger-scale model variants.

\subsection{Comparison with deep speaker embeddings}

To further contextualise the speech SSL representations, five variants of three popular deep speaker embedding models are probed: ECAPA-TDNN (E-TDNN) ~\cite{ecapa-tdnn}, the ResNet-based $r$-vector~\cite{r-vector}, and CAM++~\cite{cam++}. These models have simpler architectures, are trained on the VoxCeleb2 dataset~\cite{voxceleb2} specifically for SV, and produce a single embedding vector. The five speaker embedding models are retrieved from \textit{WeSpeaker}~\cite{wespeaker}. The probing results are illustrated in Table~\ref{tab:embed}.

As expected, the deep speaker embedding models achieve near-ceiling probing accuracies for \textbf{Speaker} and \textbf{Gender} classification, confirming their high level of specialisation. Their probing performance on other prosodic features and paralinguistic traits, however, is lower than that of the speech SSL models. This indicates that the selected prosodic attributes are less accessible from the specialised speaker embeddings under the present probing protocol.

This comparison demonstrates that general-purpose speech SSL models provide stronger probing performance for several prosodic and paralinguistic attributes than models explicitly optimised for \textbf{Speaker} identity. Therefore, for downstream tasks requiring fine-grained analysis or control of prosody and style, intermediate SSL representations may provide more accessible attribute information than dedicated deep speaker embeddings.

\section{Limitations and Future Work}

This study is limited by the discretisation of continuous prosodic features into three levels and its reliance on TextrolSpeech annotations. This approach restricts granularity and may underestimate the richness of prosodic encoding. The utterance-level training-validation split and relatively small test set may also limit the generalisability and precision of the reported probing results. Furthermore, the use of a simple probing network presents a paradox: while low-capacity networks improve interpretability by attributing performance to easily accessible information, they may fail to uncover information encoded in more complex ways. In contrast, more powerful probing networks could reveal hidden information, but they make it harder to determine whether classification success stems from the representation itself or the network's capacity. Future work will investigate this trade-off by comparing probing networks of varying complexities. Additionally, alternative datasets and fine-grained or regression-based probing methods could be explored. Developing more refined feature categorisations to address overlapping features, such as \textbf{Pitch} and \textbf{Tempo}, is another critical direction for future research.

\section{Conclusions}

To advance the explainability of speech SSL models, this study conducted a large-scale probing analysis across multiple model families and variants. This work maps how these models hierarchically learn and distribute different types of speaker-related information. By characterising the layer-wise encoding of speaker-specific attributes across various model scales, this research translates hidden states into interpretable insights. While the observed information flow exhibits several patterns consistent with prior layer-wise analyses, the unexpected recovery of speaker-specific information in the final layers of certain speech SSL models is identified and analysed. This explainability analysis provides practical and transparent guidelines for selecting the most appropriate layer representations from speech SSL models for specialised speaker-related downstream tasks.

\bibliographystyle{IEEEtran}

\bibliography{main}

% \begin{thebibliography}{9}
% \bibitem[1]{Davis80-COP}
%   S.\ B.\ Davis and P.\ Mermelstein,
%   ``Comparison of parametric representation for monosyllabic word recognition in continuously spoken sentences,''
%   \textit{IEEE Transactions on Acoustics, Speech and Signal Processing}, vol.~28, no.~4, pp.~357--366, 1980.
% \bibitem[2]{Rabiner89-ATO}
%   L.\ R.\ Rabiner,
%   ``A tutorial on hidden Markov models and selected applications in speech recognition,''
%   \textit{Proceedings of the IEEE}, vol.~77, no.~2, pp.~257-286, 1989.
% \bibitem[3]{Hastie09-TEO}
%   T.\ Hastie, R.\ Tibshirani, and J.\ Friedman,
%   \textit{The Elements of Statistical Learning -- Data Mining, Inference, and Prediction}.
%   New York: Springer, 2009.
% \bibitem[4]{YourName17-XXX}
%   F.\ Lastname1, F.\ Lastname2, and F.\ Lastname3,
%   ``Title of your ISCSLP 2026 publication,''
%   in \textit{ISCSLP 2026 -- 23\textsuperscript{rd} Annual Conference of the International Speech Communication Association, September 18-22, Incheon, Korea, Proceedings, Proceedings}, 2026, pp.~100--104.
% \end{thebibliography}

\end{document}